\documentclass[final,authoryear,3p,times,twocolumn]{elsarticle}

\usepackage{graphicx}
\usepackage{lineno}

\usepackage{amssymb}

\usepackage{natbib}

\journal{Planetary and Space Science}

\begin{document}

\widowpenalty=10000
\clubpenalty=10000

\newcommand{\aap}{Astron. Astrophys.}

\begin{frontmatter}


\ead{hanus.home@gmail.com}
\cortext[cor1]{Corresponding author. Tel: +420 221912572. Fax: +420 221912577.}


\title{The potential of sparse photometric data in asteroid shape modeling}

\author{J. Hanu\v s\corref{cor1}}
\author{J. \v Durech}

\address{Astronomical Institute, Faculty of Mathematics and Physics,
Charles University, V Hole\v sovi\v ck\' ach 2, 180 00 Prague, Czech Republic.}

\begin{abstract}
We investigate the potential of the sparse data produced by the Catalina Sky Survey astrometric project (CSS for short) in asteroid shape and rotational state determination by the lightcurve inversion method. We show that although the photometric quality of the CSS data, compared to the dense data, is significantly worse, it is in principle possible that these data are for some asteroids with high lightcurve amplitudes sufficient for a unique shape determination. CSS data are available for $\sim 180$ asteroids for which shape models were previously derived from different photometric data sets. For 13 asteroids from this sample, we derive their unique shape models based only on CSS data, compare the two independent shape models together and discuss the reliability of models derived from only CSS data. We also use CSS data to determine shape models for asteroids with already known rotational period values, derive 12 unique models and compare previously published periods with periods determined from the full 3D modeling by the lightcurve inversion method. Finally, we test different shape resolutions used in the lightcurve inversion method in order to find reliable asteroid models.
\end{abstract}


\end{frontmatter}


\section{Introduction}

Orbital parameters are currently known for more than 400\,000 asteroids. On the other hand, rotational states and shapes were determined only for a small fraction of them. In the Minor Planet Lightcurve Database\footnote{\texttt{http://cfa-www.harvard.edu/iau/lists/Lightcurve\-Dat.html}} \citep{Warner2009}, periods for $\sim3\,500$ asteroids are stored. Convex shapes with spin axis directions were derived only for $\sim200$ asteroids, these three-dimensional models of asteroids are available in the Database of Asteroid Models from Inversion Techniques\footnote{\texttt{http://astro.troja.mff.cuni.cz/projects/aste\-roids\-3D}} \citep[DAMIT, ][]{Durech2010} maintained by the Astronomical Institute of the Charles University in Prague, Czech Republic.

All asteroid models in DAMIT were derived by the lightcurve inversion method (LI). This gradient-based method is a powerful tool that allows us to derive
basic physical properties of asteroids (the rotational state and the shape) from their disk-integrated photometry
\citep[see ][]{Kaasalainen2001a,Kaasalainen2001b, Kaasalainen2002a}. LI, in a form we use, leads to a convex shape approximation of a real asteroid (close to its convex hull) that has to be a single body (or the possible moon has to be so small that its photometric signature is indistinguishable from the noise) in a relaxed rotational state (i.e., rotates around the axis with a maximum momentum of inertia). Sidereal rotational period and the direction of the spin axis are determined simultaneously with the shape. Typical number of free parameters is $\sim$50--80.

We use two different types of disk-integrated photometry: (i) \textit{dense-in-time}, which typically consists of tens to a few
hundreds of individual data points observed during one revolution (typically several hours), and (ii) \textit{sparse-in-time}, where the typical separation of individual measurements
is large compared to the rotation period. For sparse data, we usually have a few measurements per night, such as in the case of astrometric sky surveys. Based on the type of the photometry, we use the terms ``dense lightcurves'' and ``sparse lightcurves''. Unique asteroid shape models are typically feasible if we have dense lightcurves from at least three apparition, combined dense and sparse data give us more variability.

Our long-term strategy is to enlarge the number of asteroids with known shapes and rotational states, because it can help us in understanding the physical processes that take place in the asteroid's populations, such as Near Earth Asteroids (NEAs), Main Belt Asteroids (MBAs) or even asteroids in individual families \citep[e.g., Koronis family, ][]{Slivan2002, Slivan2003}. Current distribution of periods and spin axes is the direct result of the evolution of these objects starting with their formation until the present time (several hundreds of Myr to $\sim 4$ Gyr for most studied asteroids). The most prominent processes acting on asteroids are collisions and mass shedding, and the YORP effect\footnote{Yarkovsky--O'Keefe--Radzievskii--Paddack effect, a torque caused
by the recoil force from anisotropic thermal emission which can alter rotational periods and orientation of spin axes} \citep{Rubincam2000,Vokrouhlicky2003}. The knowledge of the shape can be used for several purposes, e.g.:
(i) in the construction of a thermal model \citep[e.g., ][]{Muller2011}, where values for geometric albedo, size and surface properties can be determined,
(ii) a sample of real shapes instead of synthetic ones can be used for the statistical study of the non-gravitational forces (Yarkovsky\footnote{a thermal force acting on a rotating asteroid which can alter semi-major axes} and YORP effects), or
(iii) in combination with star occultations by asteroids. These events (observed for hundreds of asteroids) give us additional information about the shape (e.g. non-convexities) and can lead to a size estimation with a typical uncertainty of 10\% \citep[see ][]{Durech2011}.

Most of the currently available photometric data were already used in the LI. The only significant exception are data from the Catalina Sky Survey astrometric project \citep[CSS for short, ][]{Larson2003}. In this paper, we investigate the possibility of determination of \textit{reliable} asteroid shape models from only a small amount of sparse-in-time distributed low-quality photometric measurements ($\sim100$) by the LI. Such data from CSS project are available for several thousands of asteroids. We want to find out if these data are for asteroids with high lightcurve amplitudes of a sufficient amount and quality for a unique shape determination, and if so, how reliable these asteroid models are. The investigation of sparse data capabilities and the reliability of derived models is important, because
(i) it can lead to a determination of new asteroid models without a need of observing any additional photometric lightcurves, and
(ii) in a few years, another huge amount of sparse data from three astrometric surveys will be available -- from the Pan-STARRS \citep[Panoramic Survey Telescope and Rapid Response System, ][]{Hodapp2004}, and later also from the Gaia satellite \citep{Perryman2001}, and the LSST \citep[Large Synoptic Survey Telescope, ][]{Ivezic2008}. Understanding the CSS data with respect to the asteroid shape modeling will speed up future processing and use of the new sparse data.

\section{Photometry}

For a unique\footnote{We define a unique solution as follows: (i) the best period has at least 10\% lower $\chi^2$ than all other periods in the scanned interval, (ii) for this period, there is only one pole solution with at least 10\% lower $\chi^2$ than the others (with a possible ambiguity for $\lambda \pm 180^{\circ}$), and (iii) this solution fulfills all our additional tests \citep[discussed in ][]{Hanus2011}.} shape determination of an asteroid, we typically need \textit{dense} photometric data from at least three apparitions and a good coverage of the solar phase angle\footnote{the Sun--asteroid--Earth angle} (i.e., different viewing geometries), $\sim20$ such dense lightcurves are usually sufficient. The biggest limiting factor of these data is their insufficient amount. Dense lightcurves were observed for $\sim3\,500$ asteroids. Only in $\sim100$ cases, asteroid shape models were derived. Determination of new models based on the dense data is now possible only if new lightcurves are observed. Many amateurs or semi-professionals use their telescopes for asteroid observations, but their main goal is the determination of the synodic period, which can be derived already from one apparition. If a period for a particular asteroid is then secure, they usually do not observe this object any more. We have to observe such object by ourselves. This approach is demanding on observational time and also on fundings. Models based only on dense data are important in testing the reliability of models based on sparse data (due to possible comparison).

Currently available \textit{sparse} photometry is accessible via the Asteroids -- Dynamic Site database\footnote{\texttt{http://hamilton.dm.unipi.it/astdys/}} (AstDyS), where data from hundreds of astrometric observatories are stored. The photometry is mostly a by-product of astrometric measurements and in most cases, asteroid magnitudes are given to only one decimal place, so the accuracy is 0.1 mag at best. Whether or nor this is sufficient for a unique shape determination for a reasonable number of asteroids was studied in \citet{Hanus2011}. The authors have found 7 observatories with quality data and used these data in combination with relative lightcurves for asteroid shape modeling. The most accurate sparse photometry is from the U.S. Naval Observatory in Flagstaff (USNO-Flagstaff station, MPC code 689). These data were already studied in \citet{Durech2009} and based on these data, the authors derived convex shape models for 24 asteroids. The biggest disadvantage of the USNO-Flagstaff station data is that they are available only for about 1000 brightest asteroids. Data from Hipparcos satellite  \citep{ESA1997} and Roque de los Muchachos Observatory, La Palma (MPC code 950) are also accurate (compared to the other observatories), but their amount of measurements is even smaller than for the USNO-Flagstaff station. The largest amount of sparse photometry is available from the Catalina Sky Survey observatory \citep[CSS for short, ][]{Larson2003}. These data were already used in the LI by \citet{Hanus2011}, but only in combination with other photometric data. Their typical photometric accuracy was $\sim12$\%. Although this accuracy of the CSS photometry seems very low, the data are valuable for asteroids with lightcurve amplitudes higher than $\gtrsim0.3$ mag, which is about a half of all asteroids \citep[based on the lightcurve data of the Minor Planet Lightcurve Database, ][]{Warner2009}. There are dense lightcurves for only $\sim700$ asteroids with numbers over ten thousand, on the other hand, we have at least 100 sparse data points from CSS for $\sim6\,000$ of these asteroids.

\section{Reliability of asteroid models derived only from the Catalina Sky Survey photometric data}

The Catalina Sky Survey astrometric project has produced the largest amount of sparse photometric measurements. For most asteroids, no other data of sufficient quality are available.

\paragraph{Test 1}
One way how to test the reliability of models based only on CSS data is to compare them with models derived from different photometric data set. So far, about a hundred of asteroid models were derived from dense data \citep[e.g., ][]{Kaasalainen2002a, Torppa2003, Slivan2003, Marciniak2007, Durech2007}, another $\sim100$ models were derived from combined dense and sparse data \citep{Durech2009, Hanus2011}. These models are believed to be reliable and thus \textit{correct}. For most of these asteroids, CSS data are also available, so we tried to derive their models from only CSS data itself. We had at least 50 individual data points for 185 asteroids. In 13 cases, we were able to derive their unique solutions. Only in 7 cases the solution was in an agreement with the model based on a different data set, other six solutions were formally correct, but derived periods were different from the expected ones.

\paragraph{Test 2}
The main inconvenient when computing models from sparse data is that synodic periods are known only for $\sim3\,500$ asteroids (determined from dense lightcurves and stored in the Minor Planet Lightcurve Database). This a priori information is used in the shape modeling: we search for the model only near the assumed period (e.g., we test the initial period values from an interval ($0.95\,P$, $1.05\,P$), where $P$ is the published period value). Typically, we test a huge number of initial period values (usually thousands), because the modeled parameter space is full of local minims and the gradient-based LI method converges to the nearest local minimum corresponding to the initial period value. The minimal difference between two local minims is given mainly by the timespan of the observational data (see Eq.~\ref{deltaP}) and is usually very small ($\lesssim10^{-3}$ h). Having a short period interval saves a lot of computational time and gives us confidence that the model is correct if the derived period value is close to $P$. Unfortunately, for most asteroids, we do not know the period (cannot be directly determined from sparse data without the model computation as in the case of the dense data), so we have to search for the model on a period interval of all possible period values, typically 2--100 hours. This consumes a lot of computational time and we loose the possibility of comparing the two period values.

\begin{figure*}[t]
\begin{center}
\includegraphics[scale=0.6]{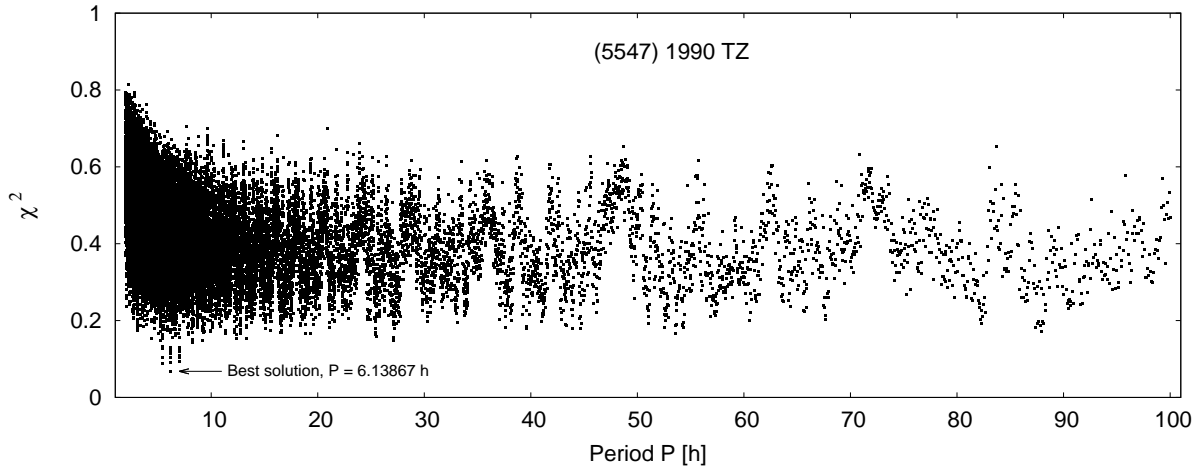}
\end{center}
\caption{\label{5647}Periodogram of asteroid (5647) 1990 TZ. Each point corresponds to a particular local minimum in the modeled parameter space of periods, pole directions and shapes.}
\end{figure*}

In this test we used the a priori information about the periods: for some asteroids, we knew their periods, we had only CSS data (there exists their dense lightcurves, but we do not have them in an electronic form), and, based on that data, we were also able to derive their models near the published period. We extended the model search on the period interval of 2--100 hours and tried to find the model solutions again (as would be done if the period was unknown). This was performed for 12 asteroids. In 7 cases, the previous solution derived on a shorter period interval was reproduced also on a larger period interval. For 4 asteroids, we did not get a unique period solution on the extended period interval, and a formally correct (i.e., it fulfilled our conditions for a unique solution), but a different solution was found for one asteroid. On Fig. \ref{5647}, we show a periodogram of asteroid (5647) 1990 TZ, where for each initial period value a $\chi^2$-value corresponding to the best shape model and pole direction (a local minimum in the multi-dimensional parameter space) is plotted. This model computation was based on 87 individual measurements from CSS, derived period $P = 6.13867$ h is in agreement with a period $P = 6.141$~h reported by \citet{Bembrick2003}.

\paragraph{Test 3}
Since we derived several unique models that were clearly incorrect, we performed this test to check more carefully the stability of the solution and to detect false solutions. The shape of an asteroid is represented by coefficients of its expansion into the spherical harmonic functions. These coefficients are optimized in the LI. The number of coefficients is given by the order $n$ of the shape expansion we use, we call $n$ the shape \textit{resolution}. In the LI, we typically use $n=6$, which corresponds to 49 coefficients ($(n+1)^2$). For asteroids with $\lesssim100$ sparse data points from CSS, so many coefficients could be to much and the LI could then lead to a random unstable solution that is even unique and physically correct. Another input parameter of the LI is the period step $P_{\mathrm{step}}$ between two subsequent periods from the scanned interval. The minimal difference between two local minims in the modeled parameter space of periods is given by

\begin{equation}\label{deltaP}
 \Delta P = \frac{P^2}{T},
\end{equation}
where $P$ is the period value and $T$ the timespan of the observational data. In our computations, we use for the period step a value $P_{\mathrm{step}}=0.8\Delta P$.

We tested the stability of the solution on different values of the resolution $n$ and on two values of the period step $P_{\mathrm{step}}$ ($0.8\Delta P$ and $0.5\Delta P$). Table \ref{resolution_summary} summarizes the results of this test for all asteroids in \textit{Test 1} for which we derived their unique solutions for $n=6$ and $P_{\mathrm{step}}=0.8\Delta P$ (corresponds to column 6). Correct unique solution is marked by ``\checkmark'', wrong unique solution by ``x'' and no unique solution by ``--''.

\begin{table*}\caption{\label{resolution_summary}Test of the stability of the model solution on the shape resolution $n$ and the step in the period $P_{\mathrm{step}}$. First column gives the asteroid number, symbol ``\checkmark'' means that for a given $n$ and $P_{\mathrm{step}}$, both unique and \textit{correct} solution was derived, symbol ``x'' means that a unique but \textit{wrong} solution was derived, and finally, in the case if no unique solution was determined we use the symbol ``--''.}
\begin{center}
 \begin{tabular}{c|ccccc|ccccc}\hline
   & \multicolumn{5}{c|} {$P_{\mathrm{step}}=0.8\Delta P$} & \multicolumn{5}{c} {$P_{\mathrm{step}}=	0.5\Delta P$}\\ \hline
 Ast & n=2 & n=3 & n=4 & n=5 & n=6 & n=2 & n=3 & n=4 & n=5 & n=6\\ \hline\hline
 685   & \checkmark & \checkmark & -- 	      & \checkmark & \checkmark & \checkmark & \checkmark & --	       & --	    & \checkmark \\
 1022  & \checkmark & \checkmark & -- 	      & --	   & \checkmark & \checkmark & \checkmark & --	       & --	    & \checkmark \\
 1419  & \checkmark & \checkmark & \checkmark & \checkmark & \checkmark & \checkmark & \checkmark & \checkmark & \checkmark & \checkmark \\
 1568  & \checkmark & \checkmark & \checkmark & \checkmark & \checkmark & \checkmark & \checkmark & \checkmark & \checkmark & \checkmark \\
 2156  & \checkmark & \checkmark & \checkmark & \checkmark & \checkmark & \checkmark & \checkmark & \checkmark & \checkmark & \checkmark \\
 3678  & -- 	    & -- 	 & \checkmark & \checkmark & \checkmark & --	     & --	  & \checkmark & --	    & \checkmark \\
 4483  & \checkmark & \checkmark & \checkmark & \checkmark & \checkmark & \checkmark & \checkmark & \checkmark & \checkmark & \checkmark \\ \hline
 52    & -- 	    & -- 	 & x 	      & -- 	   & x	        & --	     & --	  & x	       & --	    & --	 \\
 68    & -- 	    & -- 	 & -- 	      & -- 	   & x	        & --	     & --	  & --	       & --	    & x		 \\
 216   & --	    & -- 	 & -- 	      & x 	   & x	        & --	     & --	  & --	       & --	    & --	 \\
 455   & -- 	    & -- 	 & x 	      & x 	   & x	        & --	     & --	  & x	       & x	    & x		 \\
 984   & -- 	    & -- 	 & -- 	      & -- 	   & x	        & --	     & --	  & --	       & x	    & --	 \\
 1382  & -- 	    & -- 	 & -- 	      & -- 	   & x	        & --	     & --	  & --	       & --	    & --	 \\
 \end{tabular} 
\end{center}
\end{table*}

\section{Discussion}

In \textit{Test 1}, where we compared the derived models with corresponding models based on different data sets (stored in DAMIT), in 7 cases out of 13, the solution was in a well agreement with a model based on a different and higher data set. From this point of view, CSS data, which represents the largest dataset of sparse photometry with photometric accuracy theoretically sufficient for a unique shape determination, seem to be promising for the asteroid shape modeling. Unfortunately, several clearly wrong models that fulfilled our conditions on a unique solution were derived (they had different sidereal rotational periods than was expected). Similar problem occurred in \textit{Test 2}, where we compared the derived models with their previously known periods. One model was clearly incorrect.

There are several possible explanations why we derived incorrect models from the CSS data:
(i)~data for asteroids with low amplitudes are more noisy,
(ii)~low amount of data with respect to the number of modeled parameters,
(iii)~systematic errors of the survey.
In cases (i) and (ii), the data are poor, contain not enough information about the period and shape, and could produce a random solution with an unrealistic rms value (the fit is ``too perfect''). \citet{Oszkiewicz2011} show that photometric data sets from astrometric stations (such as CSS) exhibit magnitude variations with apparent V-band magnitude. For the brightest asteroids, the images are saturated, and for the faintest, the background subtraction is imperfect. This consequence of (iii) is probably partially visible in our data -- most of the incorrect models are brighter (have higher absolute magnitudes) than the correct ones. We are not able to distinguish between the effects of the amplitude and the noise from each other only from the sparse data itself without a model computation, so there is no way how to detect the case (i). The amount of data sufficient for a correct unique shape determination is dependent mainly on the time epochs (geometry of observations) and also on the signal-to-noise ratio, which we do not know. So, the sufficient amount of sparse measurements strongly varies from one asteroid to the other.

As seen from Table \ref{resolution_summary}, models derived correctly for resolution $n=6$ and period step $P_{\mathrm{step}}=0.8\Delta P$ are mostly determined for other resolutions $n$ and both period steps $P_{\mathrm{step}}$. No wrong solution appeared. In several cases, a unique solution was not found, but the correct period was still detected (we had multiple pole solutions). On the other hand, for incorrect models from \textit{Test 1}, we got unique solutions for different resolutions $n$ and period steps $P_{\mathrm{step}}$ only rarely and never for $n=2$ or $n=3$. All these unique solutions were always also incorrect. There is a weak dependence on the period step, $P_{\mathrm{step}}=0.5\Delta P$ seems to be more convenient because when used, less incorrect models were derived. To minimize the determination of incorrect models that satisfy our requirement on unique solutions, we should use the values $n=6$ and $P_{\mathrm{step}}=0.5\Delta P$ for resolution and period step, respectively. If a unique solution is then determined, the computation should be rerun with $n=3$ and $P_{\mathrm{step}}=0.5\Delta P$. The model is then reliable if we get similar unique solutions.

We rerun the computation for all 8 successfully derived models from \textit{Test 2} with $n=3$ and $P_{\mathrm{step}}=0.5\Delta P$. Previously correct solutions were confirmed, the incorrect one was ruled out.

Our method assumes a single body in a relaxed rotation state (rotates around the principal axis with the maximum momentum of inertia). Objects like tumblers or binary asteroids are not identified by the used form of the LI, their model determination usually fails because the single period model does not cover the more complex period features. The only exception are fully synchronous close binary systems with similar-sized components, such as 91 Antiope \citep{Merline2000}. These objects are reproduced as highly elongated single bodies, and because they have typically high lightcurve amplitudes, they could be easily present in the possible sample of models based on CSS data.

\section{Conclusions}

Sparse data from the Catalina Sky Survey are low efficient in unique model determination and seem more accurate for less brighter asteroids. Incorrectly determined models could be indicated by unrealistic rms values, but not generally.

Correct period solutions for a single asteroid but different resolution $n$ were always stable, the dispersion in the period was typically lower than $\Delta P$ (given by Eq.~\ref{deltaP}), which is for a period value of 10 hours and the timespan of the observations 10 years $\sim10^{-3}$ h (typical values for the period and the timespan of the CSS data). The dispersion in the ecliptic latitude of the pole direction was usually $\lesssim20^{\circ}$, but in the ecliptic longitude of the pole direction often $\gtrsim50^{\circ}$. This significant dispersion in longitude was partially caused by models with high values of ecliptic latitudes (longitudes are more dense near the pole than near the equator) and because of the small \textit{resolution} of the shapes.

We showed that CSS data can be successfully used for reliable asteroid model determination. The most accurately derived parameter is the rotational period. Although the Gaia satellite will produce during its operational time in average only $\sim60$ data points for each asteroid (for CSS, we have $\sim100$), the photometric accuracy will be incomparably better than for CSS data. Less data will be compensated by better accuracy allowing determination of many new asteroid models (or at least accurate values for rotational periods). The combination of Gaia data with data from Pan-STARRS and LSST projects will significantly increase the amount of new models. For asteroids with low amplitudes, their signal-to-noise ratio of the Gaia data will be also low and we will encounter the same problem of incorrect solutions as for CSS data. The detection of these false models will be similar.

Finding an infallible test for detecting the false solutions derived from Catalina data could allow us determination of shapes and physical parameters for tens of asteroids. While the LI is more successful for asteroids with higher lightcurve amplitudes, the derived sample of asteroid parameters will be significantly biased. For example, shapes will be more elongated than it is usual in the whole population. Careful de-biasing of the model sample will be necessary before its use in any statistical study and physical interpretation.

Possible models based on Catalina data are good candidates for follow-up observations, new dense photometry should help in the detection of incorrect solutions and should lead to more detailed shape models.

\section*{Acknowledgements}
The work of JH and JD has been supported by grant GA\,UK 134710 of the Grant agency of the Charles University, by the project SVV 265301 of the Charles University in Prague, by grants GACR 205/08/H005 and GACR 209/10/0537 of the Czech grant agency and by the Research Program MSM0021620860 of the Czech Ministry of Education. The calculations were performed on the computational cluster Tiger at the Astronomical Institute of the Charles University in Prague (\texttt{http://sirrah.troja.mff.cuni.cz/tiger}).

\bibliographystyle{model2-names}
\bibliography{mybib}

\begin{thebibliography}{24}
\expandafter\ifx\csname natexlab\endcsname\relax\def\natexlab#1{#1}\fi
\expandafter\ifx\csname url\endcsname\relax
  \def\url#1{\texttt{#1}}\fi
\expandafter\ifx\csname urlprefix\endcsname\relax\def\urlprefix{URL }\fi
\providecommand{\eprint}[2][]{\url{#2}}
\providecommand{\bibinfo}[2]{#2}
\ifx\xfnm\relax \def\xfnm[#1]{\unskip,\space#1}\fi
\bibitem[{{Bembrick} and {Bolt}(2003)}]{Bembrick2003}
\bibinfo{author}{{Bembrick}, C.}, \bibinfo{author}{{Bolt}, G.},
  \bibinfo{year}{2003}.
\newblock \bibinfo{title}{{Lightcurves across Australia: period determination
  for minor planet (5647) 1990 TZ}}.
\newblock \bibinfo{journal}{Minor Planet Bulletin} \bibinfo{volume}{30},
  \bibinfo{pages}{42--43}.
\bibitem[{{\v{D}urech} et~al.(2011){\v{D}urech}, {Kaasalainen}, {Herald},
  {Dunham}, {Timerson}, {Hanu\v{s}}, {Frappa}, {Talbot}, {Hayamizu}, {Warner},
  {Pilcher} and {Gal{\'a}d}}]{Durech2011}
\bibinfo{author}{{\v{D}urech}, J.}, \bibinfo{author}{{Kaasalainen}, M.},
  \bibinfo{author}{{Herald}, D.}, \bibinfo{author}{{Dunham}, D.},
  \bibinfo{author}{{Timerson}, B.}, \bibinfo{author}{{Hanu\v{s}}, J.},
  \bibinfo{author}{{Frappa}, E.}, \bibinfo{author}{{Talbot}, J.},
  \bibinfo{author}{{Hayamizu}, T.}, \bibinfo{author}{{Warner}, B.D.},
  \bibinfo{author}{{Pilcher}, F.}, \bibinfo{author}{{Gal{\'a}d}, A.},
  \bibinfo{year}{2011}.
\newblock \bibinfo{title}{{Combining asteroid models derived by lightcurve
  inversion with asteroidal occultation silhouettes}}.
\newblock \bibinfo{journal}{Icarus} \bibinfo{volume}{214},
  \bibinfo{pages}{652--670}.
\newblock \eprint{1104.4227}.
\bibitem[{{\v Durech} et~al.(2009){\v Durech}, {Kaasalainen}, {Warner},
  {Fauerbach}, {Marks}, {Fauvaud}, {Fauvaud}, {Vugnon}, {Pilcher}, {Bernasconi}
  and {Behrend}}]{Durech2009}
\bibinfo{author}{{\v Durech}, J.}, \bibinfo{author}{{Kaasalainen}, M.},
  \bibinfo{author}{{Warner}, B.D.}, \bibinfo{author}{{Fauerbach}, M.},
  \bibinfo{author}{{Marks}, S.A.}, \bibinfo{author}{{Fauvaud}, S.},
  \bibinfo{author}{{Fauvaud}, M.}, \bibinfo{author}{{Vugnon}, J.},
  \bibinfo{author}{{Pilcher}, F.}, \bibinfo{author}{{Bernasconi}, L.},
  \bibinfo{author}{{Behrend}, R.}, \bibinfo{year}{2009}.
\newblock \bibinfo{title}{{Asteroid models from combined sparse and dense
  photometric data}}.
\newblock \bibinfo{journal}{\aap} \bibinfo{volume}{493},
  \bibinfo{pages}{291--297}.
\bibitem[{{\v Durech} et~al.(2007){\v Durech}, {Scheirich}, {Kaasalainen},
  {Grav}, {Jedicke} and {Denneau}}]{Durech2007}
\bibinfo{author}{{\v Durech}, J.}, \bibinfo{author}{{Scheirich}, P.},
  \bibinfo{author}{{Kaasalainen}, M.}, \bibinfo{author}{{Grav}, T.},
  \bibinfo{author}{{Jedicke}, R.}, \bibinfo{author}{{Denneau}, L.},
  \bibinfo{year}{2007}.
\newblock \bibinfo{title}{{Physical models of asteroids from sparse photometric
  data}}, in: \bibinfo{editor}{{G.~B.~Valsecchi, D.~Vokrouhlick{\'y}, \&
  A.~Milani}} (Ed.), \bibinfo{booktitle}{{IAU Symposium}}, pp.
  \bibinfo{pages}{191--200}.
\bibitem[{{\v Durech} et~al.(2010){\v Durech}, {Sidorin} and
  {Kaasalainen}}]{Durech2010}
\bibinfo{author}{{\v Durech}, J.}, \bibinfo{author}{{Sidorin}, V.},
  \bibinfo{author}{{Kaasalainen}, M.}, \bibinfo{year}{2010}.
\newblock \bibinfo{title}{{DAMIT: a database of asteroid models}}.
\newblock \bibinfo{journal}{\aap} \bibinfo{volume}{513}, \bibinfo{pages}{A46}.
\bibitem[{{ESA}(1997)}]{ESA1997}
\bibinfo{author}{{ESA}}, \bibinfo{year}{1997}.
\newblock \bibinfo{title}{{The Hipparcos and Tycho Catalogues (ESA 1997)}}.
\newblock \bibinfo{journal}{VizieR Online Data Catalog} \bibinfo{volume}{1239},
  \bibinfo{pages}{0}.
\bibitem[{{Hanu\v{s}} et~al.(2011){Hanu\v{s}}, {\v{D}urech}, {Bro\v{z}},
  {Warner}, {Pilcher}, {Stephens}, {Oey}, {Bernasconi}, {Casulli}, {Behrend},
  {Polishook}, {Henych}, {Lehk{\'y}}, {Yoshida} and {Ito}}]{Hanus2011}
\bibinfo{author}{{Hanu\v{s}}, J.}, \bibinfo{author}{{\v{D}urech}, J.},
  \bibinfo{author}{{Bro\v{z}}, M.}, \bibinfo{author}{{Warner}, B.D.},
  \bibinfo{author}{{Pilcher}, F.}, \bibinfo{author}{{Stephens}, R.},
  \bibinfo{author}{{Oey}, J.}, \bibinfo{author}{{Bernasconi}, L.},
  \bibinfo{author}{{Casulli}, S.}, \bibinfo{author}{{Behrend}, R.},
  \bibinfo{author}{{Polishook}, D.}, \bibinfo{author}{{Henych}, T.},
  \bibinfo{author}{{Lehk{\'y}}, M.}, \bibinfo{author}{{Yoshida}, F.},
  \bibinfo{author}{{Ito}, T.}, \bibinfo{year}{2011}.
\newblock \bibinfo{title}{{A study of asteroid pole-latitude distribution based
  on an extended set of shape models derived by the lightcurve inversion
  method}}.
\newblock \bibinfo{journal}{\aap} \bibinfo{volume}{530}, \bibinfo{pages}{A134}.
\newblock \eprint{1104.4114}.
\bibitem[{{Hodapp} et~al.(2004){Hodapp}, {Kaiser}, {Aussel}, {Burgett},
  {Chambers}, {Chun}, {Dombeck}, {Douglas}, {Hafner}, {Heasley}, {Hoblitt},
  {Hude} and {\textit{et al.}}}]{Hodapp2004}
\bibinfo{author}{{Hodapp}, K.W.}, \bibinfo{author}{{Kaiser}, N.},
  \bibinfo{author}{{Aussel}, H.}, \bibinfo{author}{{Burgett}, W.},
  \bibinfo{author}{{Chambers}, K.C.}, \bibinfo{author}{{Chun}, M.},
  \bibinfo{author}{{Dombeck}, T.}, \bibinfo{author}{{Douglas}, A.},
  \bibinfo{author}{{Hafner}, D.}, \bibinfo{author}{{Heasley}, J.},
  \bibinfo{author}{{Hoblitt}, J.}, \bibinfo{author}{{Hude}, C.},
  \bibinfo{author}{{\textit{et al.}}}, \bibinfo{year}{2004}.
\newblock \bibinfo{title}{{Design of the Pan-STARRS telescopes}}.
\newblock \bibinfo{journal}{Astronomische Nachrichten} \bibinfo{volume}{325},
  \bibinfo{pages}{636--642}.
\bibitem[{{Ivezi{\'c}} et~al.(2008){Ivezi{\'c}}, {Tyson}, {Acosta}, {Allsman},
  {Anderson}, {Andrew}, {Angel}, {Axelrod}, {Barr}, {Becker}, {Becla},
  {Beldica} and {\textit{et al.}}}]{Ivezic2008}
\bibinfo{author}{{Ivezi{\'c}}, Z.}, \bibinfo{author}{{Tyson}, J.A.},
  \bibinfo{author}{{Acosta}, E.}, \bibinfo{author}{{Allsman}, R.},
  \bibinfo{author}{{Anderson}, S.F.}, \bibinfo{author}{{Andrew}, J.},
  \bibinfo{author}{{Angel}, R.}, \bibinfo{author}{{Axelrod}, T.},
  \bibinfo{author}{{Barr}, J.D.}, \bibinfo{author}{{Becker}, A.C.},
  \bibinfo{author}{{Becla}, J.}, \bibinfo{author}{{Beldica}, C.},
  \bibinfo{author}{{\textit{et al.}}}, \bibinfo{year}{2008}.
\newblock \bibinfo{title}{{LSST: from Science Drivers to Reference Design and
  Anticipated Data Products}}.
\newblock \bibinfo{journal}{ArXiv e-prints} \eprint{0805.2366}.
\bibitem[{{Kaasalainen} et~al.(2002){Kaasalainen}, {Mottola} and
  {Fulchignoni}}]{Kaasalainen2002a}
\bibinfo{author}{{Kaasalainen}, M.}, \bibinfo{author}{{Mottola}, S.},
  \bibinfo{author}{{Fulchignoni}, M.}, \bibinfo{year}{2002}.
\newblock \bibinfo{title}{{Asteroid Models from Disk-integrated Data}}.
\newblock \bibinfo{journal}{Asteroids III} , \bibinfo{pages}{139--150}.
\bibitem[{{Kaasalainen} and {Torppa}(2001)}]{Kaasalainen2001a}
\bibinfo{author}{{Kaasalainen}, M.}, \bibinfo{author}{{Torppa}, J.},
  \bibinfo{year}{2001}.
\newblock \bibinfo{title}{{Optimization Methods for Asteroid Lightcurve
  Inversion. I. Shape Determination}}.
\newblock \bibinfo{journal}{Icarus} \bibinfo{volume}{153},
  \bibinfo{pages}{24--36}.
\bibitem[{{Kaasalainen} et~al.(2001){Kaasalainen}, {Torppa} and
  {Muinonen}}]{Kaasalainen2001b}
\bibinfo{author}{{Kaasalainen}, M.}, \bibinfo{author}{{Torppa}, J.},
  \bibinfo{author}{{Muinonen}, K.}, \bibinfo{year}{2001}.
\newblock \bibinfo{title}{{Optimization Methods for Asteroid Lightcurve
  Inversion. II. The Complete Inverse Problem}}.
\newblock \bibinfo{journal}{Icarus} \bibinfo{volume}{153},
  \bibinfo{pages}{37--51}.
\bibitem[{{Larson} et~al.(2003){Larson}, {Beshore}, {Hill}, {Christensen},
  {McLean}, {Kolar}, {McNaught} and {Garradd}}]{Larson2003}
\bibinfo{author}{{Larson}, S.}, \bibinfo{author}{{Beshore}, E.},
  \bibinfo{author}{{Hill}, R.}, \bibinfo{author}{{Christensen}, E.},
  \bibinfo{author}{{McLean}, D.}, \bibinfo{author}{{Kolar}, S.},
  \bibinfo{author}{{McNaught}, R.}, \bibinfo{author}{{Garradd}, G.},
  \bibinfo{year}{2003}.
\newblock \bibinfo{title}{{The CSS and SSS NEO surveys}}, in:
  \bibinfo{booktitle}{{AAS/Division for Planetary Sciences Meeting Abstracts
  \#35}}, p. \bibinfo{pages}{982}.
\bibitem[{{Marciniak} et~al.(2007){Marciniak}, {Micha{\l}owski}, {Kaasalainen},
  {\v{D}urech}, {Poli{\'n}ska}, {Kwiatkowski}, {Kryszczy{\'n}ska}, {Hirsch},
  {Kami{\'n}ski}, {Fagas}, {Colas}, {Fauvaud}, {Santacana}, {Behrend} and
  {Roy}}]{Marciniak2007}
\bibinfo{author}{{Marciniak}, A.}, \bibinfo{author}{{Micha{\l}owski}, T.},
  \bibinfo{author}{{Kaasalainen}, M.}, \bibinfo{author}{{\v{D}urech}, J.},
  \bibinfo{author}{{Poli{\'n}ska}, M.}, \bibinfo{author}{{Kwiatkowski}, T.},
  \bibinfo{author}{{Kryszczy{\'n}ska}, A.}, \bibinfo{author}{{Hirsch}, R.},
  \bibinfo{author}{{Kami{\'n}ski}, K.}, \bibinfo{author}{{Fagas}, M.},
  \bibinfo{author}{{Colas}, F.}, \bibinfo{author}{{Fauvaud}, S.},
  \bibinfo{author}{{Santacana}, G.}, \bibinfo{author}{{Behrend}, R.},
  \bibinfo{author}{{Roy}, R.}, \bibinfo{year}{2007}.
\newblock \bibinfo{title}{{Photometry and models of selected main belt
  asteroids. IV. 184 Dejopeja, 276 Adelheid, 556 Phyllis}}.
\newblock \bibinfo{journal}{\aap} \bibinfo{volume}{473},
  \bibinfo{pages}{633--639}.
\bibitem[{{Merline} et~al.(2000){Merline}, {Close}, {Dumas}, {Shelton},
  {Menard}, {Chapman} and {Slater}}]{Merline2000}
\bibinfo{author}{{Merline}, W.J.}, \bibinfo{author}{{Close}, L.M.},
  \bibinfo{author}{{Dumas}, C.}, \bibinfo{author}{{Shelton}, J.C.},
  \bibinfo{author}{{Menard}, F.}, \bibinfo{author}{{Chapman}, C.R.},
  \bibinfo{author}{{Slater}, D.C.}, \bibinfo{year}{2000}.
\newblock \bibinfo{title}{{Discovery of Companions to Asteroids 762 Pulcova and
  90 Antiope by Direct Imaging}}, in: \bibinfo{booktitle}{{AAS/Division for
  Planetary Sciences Meeting Abstracts \#32}}, p. \bibinfo{pages}{1017}.
\bibitem[{{M{\"u}ller} et~al.(2011){M{\"u}ller}, {\v{D}urech}, {Hasegawa},
  {Abe}, {Kawakami}, {Kasuga}, {Kinoshita}, {Kuroda}, {Urakawa}, {Okumura},
  {Sarugaku}, {Miyasaka}, {Takagi}, {Weissman}, {Choi}, {Larson}, {Yanagisawa}
  and {Nagayama}}]{Muller2011}
\bibinfo{author}{{M{\"u}ller}, T.G.}, \bibinfo{author}{{\v{D}urech}, J.},
  \bibinfo{author}{{Hasegawa}, S.}, \bibinfo{author}{{Abe}, M.},
  \bibinfo{author}{{Kawakami}, K.}, \bibinfo{author}{{Kasuga}, T.},
  \bibinfo{author}{{Kinoshita}, D.}, \bibinfo{author}{{Kuroda}, D.},
  \bibinfo{author}{{Urakawa}, S.}, \bibinfo{author}{{Okumura}, S.},
  \bibinfo{author}{{Sarugaku}, Y.}, \bibinfo{author}{{Miyasaka}, S.},
  \bibinfo{author}{{Takagi}, Y.}, \bibinfo{author}{{Weissman}, P.R.},
  \bibinfo{author}{{Choi}, Y.J.}, \bibinfo{author}{{Larson}, S.},
  \bibinfo{author}{{Yanagisawa}, K.}, \bibinfo{author}{{Nagayama}, S.},
  \bibinfo{year}{2011}.
\newblock \bibinfo{title}{{Thermo-physical properties of 162173 (1999 JU3), a
  potential flyby and rendezvous target for interplanetary missions}}.
\newblock \bibinfo{journal}{\aap} \bibinfo{volume}{525}, \bibinfo{pages}{A145}.
\newblock \eprint{1011.5029}.
\bibitem[{{Oszkiewicz} et~al.(2011){Oszkiewicz}, {Muinonen}, {Bowell},
  {Trilling}, {Penttil{\"a}}, {Pieniluoma}, {Wasserman} and
  {Enga}}]{Oszkiewicz2011}
\bibinfo{author}{{Oszkiewicz}, D.A.}, \bibinfo{author}{{Muinonen}, K.},
  \bibinfo{author}{{Bowell}, E.}, \bibinfo{author}{{Trilling}, D.},
  \bibinfo{author}{{Penttil{\"a}}, A.}, \bibinfo{author}{{Pieniluoma}, T.},
  \bibinfo{author}{{Wasserman}, L.H.}, \bibinfo{author}{{Enga}, M.T.},
  \bibinfo{year}{2011}.
\newblock \bibinfo{title}{{Online multi-parameter phase-curve fitting and
  application to a large corpus of asteroid photometric data}}.
\newblock \bibinfo{journal}{Journal of Quantitative Spectroscopy \& Radiative
  Transfer} \bibinfo{volume}{112}, \bibinfo{pages}{1919--1929}.
\bibitem[{{Perryman} et~al.(2001){Perryman}, {de Boer}, {Gilmore}, {H{\o}g},
  {Lattanzi}, {Lindegren}, {Luri}, {Mignard}, {Pace} and {de
  Zeeuw}}]{Perryman2001}
\bibinfo{author}{{Perryman}, M.A.C.}, \bibinfo{author}{{de Boer}, K.S.},
  \bibinfo{author}{{Gilmore}, G.}, \bibinfo{author}{{H{\o}g}, E.},
  \bibinfo{author}{{Lattanzi}, M.G.}, \bibinfo{author}{{Lindegren}, L.},
  \bibinfo{author}{{Luri}, X.}, \bibinfo{author}{{Mignard}, F.},
  \bibinfo{author}{{Pace}, O.}, \bibinfo{author}{{de Zeeuw}, P.T.},
  \bibinfo{year}{2001}.
\newblock \bibinfo{title}{{GAIA: Composition, formation and evolution of the
  Galaxy}}.
\newblock \bibinfo{journal}{\aap} \bibinfo{volume}{369},
  \bibinfo{pages}{339--363}.
\newblock \eprint{ArXiv:astro-ph/0101235}.
\bibitem[{{Rubincam}(2000)}]{Rubincam2000}
\bibinfo{author}{{Rubincam}, D.P.}, \bibinfo{year}{2000}.
\newblock \bibinfo{title}{{Radiative Spin-up and Spin-down of Small
  Asteroids}}.
\newblock \bibinfo{journal}{Icarus} \bibinfo{volume}{148},
  \bibinfo{pages}{2--11}.
\bibitem[{{Slivan}(2002)}]{Slivan2002}
\bibinfo{author}{{Slivan}, S.M.}, \bibinfo{year}{2002}.
\newblock \bibinfo{title}{{Spin vector alignment of Koronis family asteroids}}.
\newblock \bibinfo{journal}{Nature} \bibinfo{volume}{419},
  \bibinfo{pages}{49--51}.
\bibitem[{{Slivan} et~al.(2003){Slivan}, {Binzel}, {Crespo da Silva},
  {Kaasalainen}, {Lyndaker} and {Kr\v{c}o}}]{Slivan2003}
\bibinfo{author}{{Slivan}, S.M.}, \bibinfo{author}{{Binzel}, R.P.},
  \bibinfo{author}{{Crespo da Silva}, L.D.}, \bibinfo{author}{{Kaasalainen},
  M.}, \bibinfo{author}{{Lyndaker}, M.M.}, \bibinfo{author}{{Kr\v{c}o}, M.},
  \bibinfo{year}{2003}.
\newblock \bibinfo{title}{{Spin vectors in the Koronis family: comprehensive
  results from two independent analyses of 213 rotation lightcurves}}.
\newblock \bibinfo{journal}{Icarus} \bibinfo{volume}{162},
  \bibinfo{pages}{285--307}.
\bibitem[{{Torppa} et~al.(2003){Torppa}, {Kaasalainen}, {Micha{\l}owski},
  {Kwiatkowski}, {Kryszczy{\'n}ska}, {Denchev} and {Kowalski}}]{Torppa2003}
\bibinfo{author}{{Torppa}, J.}, \bibinfo{author}{{Kaasalainen}, M.},
  \bibinfo{author}{{Micha{\l}owski}, T.}, \bibinfo{author}{{Kwiatkowski}, T.},
  \bibinfo{author}{{Kryszczy{\'n}ska}, A.}, \bibinfo{author}{{Denchev}, P.},
  \bibinfo{author}{{Kowalski}, R.}, \bibinfo{year}{2003}.
\newblock \bibinfo{title}{{Shapes and rotational properties of thirty asteroids
  from photometric data}}.
\newblock \bibinfo{journal}{Icarus} \bibinfo{volume}{164},
  \bibinfo{pages}{346--383}.
\bibitem[{{Vokrouhlick{\'y}} et~al.(2003){Vokrouhlick{\'y}}, {Nesvorn{\'y}} and
  {Bottke}}]{Vokrouhlicky2003}
\bibinfo{author}{{Vokrouhlick{\'y}}, D.}, \bibinfo{author}{{Nesvorn{\'y}}, D.},
  \bibinfo{author}{{Bottke}, W.F.}, \bibinfo{year}{2003}.
\newblock \bibinfo{title}{{The vector alignments of asteroid spins by thermal
  torques}}.
\newblock \bibinfo{journal}{Nature} \bibinfo{volume}{425},
  \bibinfo{pages}{147--151}.
\bibitem[{{Warner} et~al.(2009){Warner}, {Harris} and {Pravec}}]{Warner2009}
\bibinfo{author}{{Warner}, B.D.}, \bibinfo{author}{{Harris}, A.W.},
  \bibinfo{author}{{Pravec}, P.}, \bibinfo{year}{2009}.
\newblock \bibinfo{title}{{The asteroid lightcurve database}}.
\newblock \bibinfo{journal}{Icarus} \bibinfo{volume}{202},
  \bibinfo{pages}{134--146}.

\end{thebibliography}

\end{document}